\documentclass[aps,pra,reprint,superscriptaddress]{revtex4-2}
\usepackage{times}
\usepackage{graphicx}
\usepackage{orcidlink}
\usepackage{amsmath,amssymb,bm}
\usepackage{physics}
\usepackage{hyperref}

\begin{document}

\title{Connectivity--Interference Competition in Coherent Transport on Percolated Hierarchical Small-World Networks}

\author{Miquéias Jacinto Cirino
\orcidlink{0000-0001-8735-0705}}
\email{[mikeyas@ifi.unicamp.br](mailto:mikeyas@ifi.unicamp.br)}
\affiliation{Instituto de Física Gleb Wataghin, Universidade Estadual de Campinas, 13083-859, Campinas, SP, Brazil}

\author{Marcos César de Oliveira
\orcidlink{0000-0003-2251-2632}}
\email{[marcos@ifi.unicamp.br](mailto:marcos@ifi.unicamp.br)}
\affiliation{Instituto de Física Gleb Wataghin, Universidade Estadual de Campinas, 13083-859, Campinas, SP, Brazil}

\date{\today}

\begin{abstract}
Adding links generally improves classical transport by increasing the number of available paths. We show that coherent quantum transport can display the opposite behavior. Using continuous-time quantum walks on a percolated hierarchical small-world network, we identify a coherent overconnectivity penalty: root-to-boundary transport is maximized at intermediate bond probability and decreases as the network approaches full connectivity. The effect is quantified by the final-layer limiting probability $\chi_N$ and by the penalty
$P_Q=1-\chi_N(p=1)/\max_p\chi_N(p)$, which measures the loss caused by making the architecture fully connected. The optimum results from a competition between shortcut-assisted spreading and interference-induced intra-layer recirculation. Spectral analysis shows that bond dilution creates motif-induced degeneracies and reorganizes the eigenstates connecting the root to the outermost layer. A comparison with dephased and classical transport shows that the non-monotonic landscape is not a purely geometrical percolation effect, but a coherent architecture-dependent phenomenon. These results provide a design principle for coherent transport in disordered photonic and quantum-network architectures.
\end{abstract}

\maketitle

Coherent propagation in a network is shaped not only by the number of available paths, but also by how those paths interfere. This makes the architecture--performance relation in quantum transport fundamentally different from that of classical diffusion. In a classical network, increasing connectivity typically improves spreading by opening additional routes. In a coherent network, however, the same additional routes may also create destructive interference, backflow, and recirculation inside highly connected regions. Thus, whether a network is well connected is not a purely geometrical question, as it depends on how the spectrum and eigenstates of the graph mediate the propagation of amplitude between input and output regions.
This issue is particularly relevant for complex quantum networks, quantum communication architectures, and photonic implementations in which disorder and broken links are unavoidable~\cite{Nokkala2024,Nikolopoulos2014,Kimble2008,Brito2020,Brito2021}. In such systems, redundant connectivity is usually desirable because it increases robustness against link failures. Yet, in a coherent setting, redundant paths also introduce additional phase-sensitive channels. The same architectural feature that creates shortcuts may therefore suppress transport if it redirects amplitude away from the target region. Identifying when connectivity acts as a resource and when it becomes detrimental is a basic design problem for coherent transport.

Continuous-time quantum walks (CTQWs) provide a minimal framework to isolate this competition because the Hamiltonian is directly determined by graph connectivity~\cite{Aharanov1993,Farhi1998,jingbook2014,Mulken2011}. Disorder in such systems is commonly studied through quantum percolation, where bond removal changes both the available paths and the spectral structure responsible for coherent propagation~\cite{Kirkpatrick1972,Shapir1982,Mookerjee1995,Schubert2005,Chandrashekar2014,Kollar2014,Chawla2019,Darazs2013,Benedetti2018}. Previous studies have shown that percolation, stochastic connectivity, and dynamical disorder can strongly modify quantum-walk spreading. Here we focus on a sharper architectural question: does maximal connectivity maximize coherent information transfer in a disordered hierarchical network?

We answer this question negatively. We consider a hierarchical small-world architecture composed of an $m$-branching tree supplemented by intra-layer small-world links. Starting from an excitation localized at the root, we show that the final-layer transport efficiency is non-monotonic in the bond probability. Coherent propagation is suppressed at low connectivity because spanning paths are rare, enhanced at intermediate connectivity by shortcut-assisted access to deeper layers, and suppressed again in the overconnected regime because intra-layer recirculation dominates root-to-boundary transfer. This produces a coherent overconnectivity penalty, as adding links beyond the optimum reduces, rather than improves, the probability of reaching the target layer.

The graph has $N$ layers, with $Q_n=m^{n-1}$ vertices in layer $n$ and total size $S_m^N=(m^N-1)/(m-1)$. Tree edges connect each vertex in layer $n$ to $m$ descendants in layer $n+1$. Each layer is also endowed with Watts--Strogatz intra-layer links controlled by a degree parameter $K$ and rewiring probability $\phi$~\cite{Watts1998}. Thus $m$ and $N$ determine the hierarchical depth and growth rate, while $K$ and $\phi$ tune the lateral connectivity responsible for shortcut formation and intra-layer redistribution. Bond percolation retains each edge with probability $p$. For each percolated realization $r$, coherent propagation is generated by the Laplacian CTQW Hamiltonian
\begin{equation}
H_r=-L_r=-(D_r-A_r),
\label{eq:hamiltonian}
\end{equation}
where $A_r$ and $D_r$ are the adjacency and degree matrices of that realization. We set the hopping-rate scale to unity. The initial state is $\ket{\psi(0)}=\ket{1}$, corresponding to the root, and observables are computed for each realization and then averaged over the ensemble. Further details of the graph construction, vertex ordering, and percolation threshold estimate are given in the Supplemental Material.

\begin{figure}[h]
\centering
\includegraphics[width=0.86\linewidth]{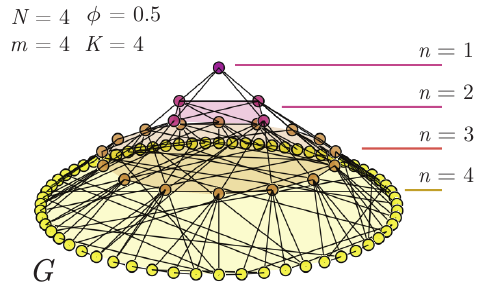}
\caption{Hierarchical small-world transport architecture. The graph combines an $m$-branching tree backbone with intra-layer small-world links. The example shown has $N=4$, $m=4$, $\phi=0.5$, and $K=4$.}
\label{fig:schematic}
\end{figure}

We emphasize that the present work addresses transport in finite hierarchical architectures rather than a thermodynamic critical-point problem. The percolation scales are used to identify geometrical regimes in which root-to-boundary paths become available, while the transport optimum is defined by the coherent limiting probability on finite networks. Thus, the central question is not the location of a universal critical threshold, but how the architecture of a disordered finite network controls the spectral pathways available for coherent propagation.

Bond percolation plays two distinct roles. Geometrically, it controls whether a connected route from the root to the final layer exists. In a pure $m$-branching tree, the corresponding bond-percolation scale is $p_c^{\rm tree}=1/m$, while intra-layer links lower the effective geometrical scale by opening lateral routes before downward propagation. Spectrally, however, percolation also changes the Hamiltonian itself: removing edges modifies both the off-diagonal hopping amplitudes and the degree-dependent diagonal terms of the Laplacian. Consequently, geometrical connectivity is necessary but not sufficient for efficient coherent transport. Even when a spanning cluster exists, root-to-boundary propagation can be suppressed if the relevant eigenstates have poor simultaneous support on the root and final-layer subspace. The transport problem is therefore not only to create paths, but to create spectral pathways that connect the input and target regions effectively.

Layer-resolved transport is quantified by the projector $\Pi_n=\sum_{i\in n}\ketbra{i}{i}$ and by the ensemble-averaged layer probability
\begin{equation}
P_n(t)=\overline{\bra{\psi_r(t)}\Pi_n\ket{\psi_r(t)}}.
\label{eq:layer_probability}
\end{equation}
The probabilities $P_1(t)$ and $P_N(t)$ respectively measure return to the root layer and transfer to the outermost layer. At small $p$, the excitation remains predominantly near the root. Increasing $p$ opens inter-layer and intra-layer pathways and increases the population reaching the final layer. At larger connectivity, the ensemble-averaged oscillations are damped by averaging over distinct coherent spectra, while each individual realization remains unitary. Representative time traces are shown in the Supplemental Material.

Since finite coherent quantum walks do not generally converge to a stationary distribution, we use the limiting probability
\begin{eqnarray}
\chi_n&=&\lim_{T\rightarrow\infty}\frac{1}{T}\int_0^T P_n(t)\,dt  \nonumber\\
&=&\overline{\sum_{i\in n}\sum_{\ell,m}\delta_{E_\ell,E_m}
\braket{i}{\varphi_\ell}\braket{\varphi_\ell}{1}
\braket{1}{\varphi_m}\braket{\varphi_m}{i}},
\label{eq:limiting}
\end{eqnarray}
where ${\ket{\varphi_\ell}}$ are eigenstates of $H_r$. The final-layer value $\chi_N$ is the root-to-boundary coherent transport efficiency. Equation~\eqref{eq:limiting} shows that transport is controlled not only by the existence of paths, but by the overlap of degenerate eigenstates with both the initial root and the target layer.

\begin{figure}[h]
\includegraphics[width=\linewidth]{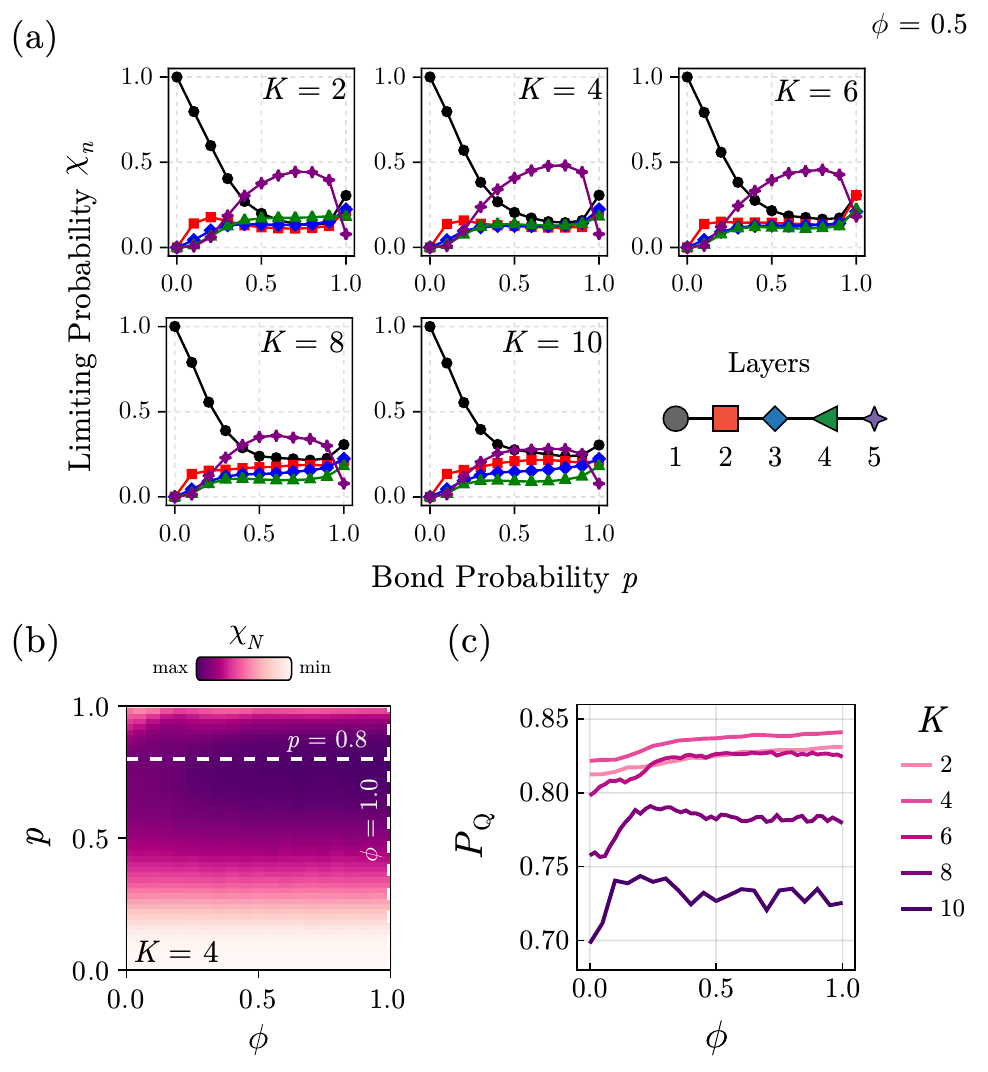}
\caption{Optimal coherent transport and overconnectivity penalty. Layer-resolved limiting probabilities $\chi_n(p)$ for several $K$ at fixed $\phi=0.5$ in (a), together with the final-layer limiting probability $\chi_N$ as a function of bond probability $p$ and rewiring probability $\phi$ for $K=4$ in (b). The maximum transfer to the final layer occurs at intermediate connectivity. (c) Overconnectivity penalty $P_Q=1-\chi_N(p=1,K,\phi)/\max_p\chi_N(p,K,\phi)$ as a function of $\phi$ for different $K$. The penalty quantifies the loss caused by making the network fully connected. All data are averaged over $R=10000$ realizations.}
\label{fig:mainresult}
\end{figure}

Figure~\ref{fig:mainresult} shows the central result. The dependence of $\chi_N$ on the bond probability naturally separates the dynamics into three regimes. At low $p$, transport is connectivity-limited: the diluted network rarely provides effective root-to-boundary paths, and the final-layer efficiency remains small. At intermediate $p$, inter-layer paths and small-world shortcuts cooperate to increase access to deeper layers, so that connectivity acts as a resource and $\chi_N$ rises rapidly. At large $p$, however, the trend reverses. The network contains more links, but dense intra-layer connectivity promotes coherent amplitude redistribution and recirculation within layers, reducing the net probability transferred to the boundary. Thus, coherent transport is not optimized by maximizing the number of links.

This non-monotonicity defines a connectivity--interference competition. Additional links provide more paths, but in a coherent system they also introduce additional interference channels and modify the eigenstate structure. The same lateral connectivity that creates shortcuts can therefore enhance intra-layer recirculation and reduce directed root-to-boundary transfer. The optimum of $\chi_N$ marks the point at which shortcut-assisted spreading and interference-induced recirculation are best balanced.

The competition also depends strongly on the intra-layer degree parameter $K$. For small $K$, lateral connectivity is insufficient to efficiently redistribute amplitude across each layer before further downward propagation. For larger $K$, lateral redistribution becomes stronger and can redirect amplitude into intra-layer pathways instead of supporting directed propagation through the hierarchy. In the parameter range studied here, $K=4$ gives the largest final-layer efficiency, revealing an optimal lateral-connectivity scale for coherent propagation.

To quantify the loss associated with maximal connectivity, we define the coherent overconnectivity penalty
\begin{equation}
P_Q(K,\phi)=1-\frac{\chi_N(p=1,K,\phi)}{\max_p\chi_N(p,K,\phi)}.
\label{eq:overconnectivity}
\end{equation}
A positive $P_Q$ means that the fully connected network is less efficient than an intermediate-connectivity architecture. This quantity converts the non-monotonicity of $\chi_N$ into a direct measure of the loss caused by overconnecting the network. As shown in Fig.~\ref{fig:mainresult}, the penalty depends strongly on $K$, indicating that the effect is not controlled by bond dilution alone. Rather, it depends on the lateral-connectivity scale that determines how strongly amplitude is redistributed within each layer before it can propagate deeper into the hierarchy. The overconnected regime therefore contains more available paths, but transports less efficiently to the target layer.

The overconnectivity penalty should therefore be read as an architecture-level diagnostic. It does not measure whether the graph is connected, but whether the fully connected architecture is spectrally efficient for transferring amplitude from the root to the boundary. A positive value of $P_Q$ indicates that adding links beyond the optimum creates coherent pathways that compete with, rather than support, directed transfer. In this sense, the fully connected graph is not a limiting ideal for coherent transport; it is a distinct dynamical regime in which lateral recirculation can dominate over root-to-boundary propagation.
Spectral analysis supports this interpretation. Bond dilution creates repeated local motifs and sharp spectral degeneracies, while intra-layer small-world links reorganize the eigenstates with support on the final layer. Increasing $K$ broadens the available spectral support, but it also redistributes eigenstate weight across intermediate and lateral subspaces. Thus, percolation determines whether paths exist, whereas coherent efficiency depends on how eigenstates distribute amplitude between the root, the intermediate layers, and the target layer. Dense lateral connectivity can therefore confine part of the coherent amplitude through intra-layer recirculation, suppressing the net transfer to the boundary. Detailed spectra and spanning-cluster overlaps are provided in the Supplemental Material.

The relevant spectral quantity is not the total number of connected
paths, but the simultaneous overlap of eigenstates with the input
state and the final-layer subspace. Increasing connectivity can enlarge
the connected component and broaden the spectrum while still reducing
the fraction of eigenstate weight that contributes efficiently to
root-to-boundary transfer. In this regime, additional links create
spectral channels that redistribute amplitude laterally rather than
enhancing the root-to-boundary overlap entering Eq.~\eqref{eq:limiting}.

\begin{figure}[t]
\centering
\includegraphics[width=\linewidth]{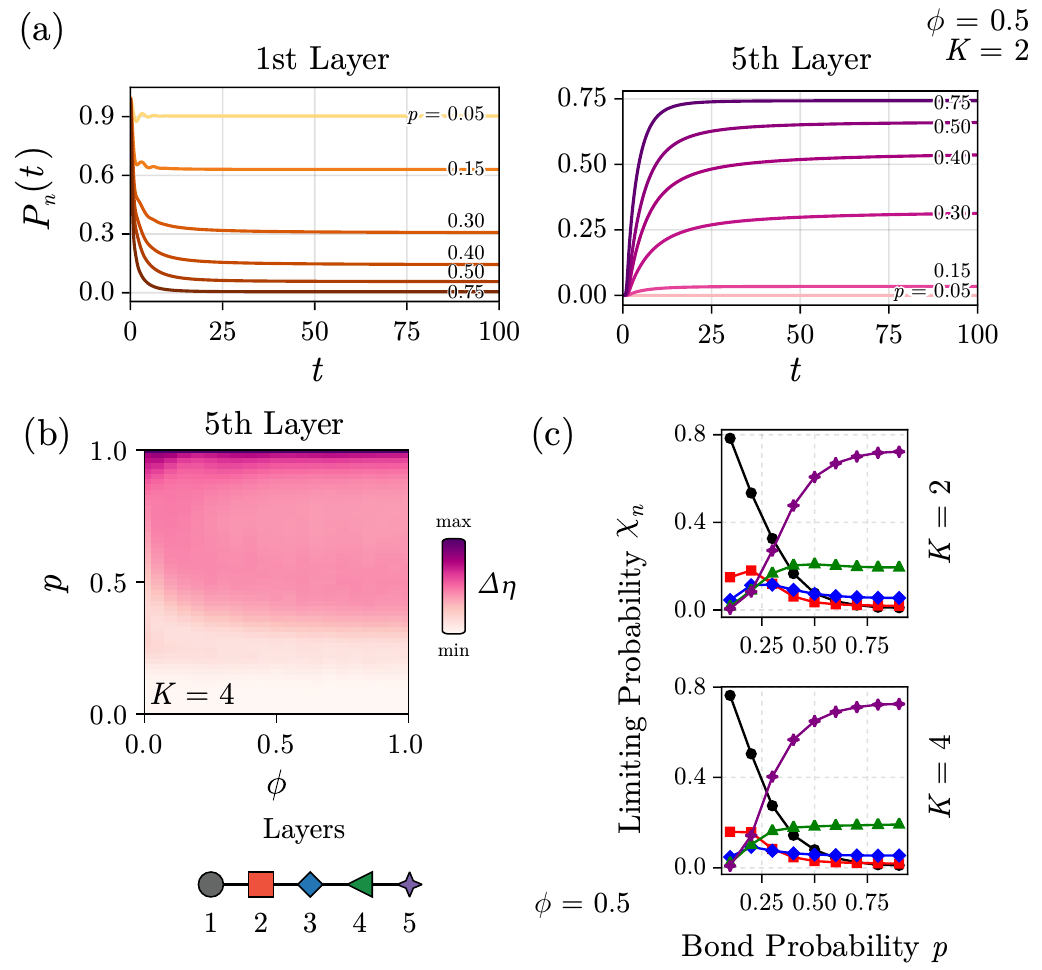}
\caption{Quantum--classical contrast. (a) Dephased layer probability for the first and final layers for fixed $(\phi,K)=(0.5,2)$ and several bond probabilities. (b) Final-layer efficiency contrast $\Delta\eta$ between the incoherent/classical benchmark and the coherent dynamics for $K=4$. (c) Layer-resolved limiting probabilities in the dephased regime. Dephasing suppresses oscillations and produces a more monotonic accumulation in the outer layers, in contrast with the coherent optimum of Fig.~\ref{fig:mainresult}.}
\label{fig:classicalcontrast}
\end{figure}

To separate coherent interference from purely geometrical effects, we compare the unitary dynamics with a site-dephased evolution,
\begin{equation}
\dot\rho_r=-i[H_r,\rho_r]
-q\left(\rho_r-\sum_j P_j\rho_r P_j\right),
\label{eq:dephasing}
\end{equation}
where $P_j=\ketbra{j}{j}$. The corresponding layer probability is $P_n^{\rm deph}(t)=\overline{\mathrm{Tr}[\Pi_n\rho_r(t)]}$. This evolution suppresses site-basis coherences while keeping the same percolated graph realization. It therefore provides an operational benchmark for distinguishing genuinely coherent transport effects from those caused only by the existence of geometrical pathways.

We quantify this difference through the final-layer contrast
$\Delta\eta=\chi_N^{\rm cl}-\chi_N$, where $\chi_N^{\rm cl}$
denotes the corresponding incoherent/classical benchmark defined in
the Supplemental Material.
Figure~\ref{fig:classicalcontrast} shows that dephasing rapidly suppresses the oscillatory component and produces a much more monotonic accumulation in the outer layers. In this regime, increasing connectivity primarily improves access to the larger final layer, and the response is governed more by coarse connectivity and layer-size distribution than by coherent spectral pathways. The final-layer contrast $\Delta\eta$ is smallest near the region where coherent transport is most efficient, but the dephased/classical response lacks the same overconnectivity optimum. Once phase-sensitive interference is removed, the detailed organization of coherent paths becomes less relevant than the coarse connectivity of the graph. The suppression of $\chi_N$ at large $p$ therefore cannot be attributed to the absence of paths; it arises from wave interference and intra-layer coherent recirculation.

We have shown that coherent transport on a percolated hierarchical small-world network exhibits an optimal connectivity window. The result can be summarized as a coherent overconnectivity penalty: increasing the number of available links beyond an optimal value reduces root-to-boundary quantum transport. This effect is absent in a purely geometrical interpretation of percolation and reflects the spectral organization of coherent propagation. In coherent architectures, topology is not merely a passive substrate for transport; it actively shapes the interference pathways through which amplitude reaches, or fails to reach, the target layer.
This distinction suggests a different notion of robustness for coherent networks. In an incoherent or classical setting, redundancy primarily increases the number of available routes. In a coherent setting, redundancy also changes the interference landscape. Robust coherent transport therefore requires not only path redundancy, but spectral compatibility between the input and output regions. The optimal network is the one in which additional links open useful shortcuts without generating excessive intra-layer recirculation.

For photonic and quantum-network architectures, the implication is that robustness should not be pursued only by adding redundant channels. Efficient coherent transport requires engineered connectivity that balances shortcut-assisted access against undesired intra-layer recirculation. Similar principles may be relevant for biologically inspired transport architectures, where hierarchical organization, structural disorder, and lateral connectivity coexist, although direct biological applications require additional physical mechanisms beyond the present tight-binding model.

\begin{acknowledgments}
This work was mainly supported by grant 2024/23699-4, S~ao Paulo Research Foundation (FAPESP). We thank the Coaraci Supercomputer for computer time (FAPESP grant 2019/17874-0). MCO acknowledges partial financial support from the National Institute of Science and Technology for Applied Quantum Computing through CNPq Process No. 408884/2024-0 and by FAPESP, through the Center for Research and Innovation on Smart and Quantum Materials (CRISQuaM) Process No. 2024/00998-6.
\end{acknowledgments}

\bibliographystyle{apsrev4-2}
\bibliography{dynamical_percolation}

\appendix

\section*{Supplemental Information and Methods}

\subsection{Network construction and ensemble averaging}

The transport architecture is an undirected graph $G=G(V,E)$ with an $m$-branching hierarchical backbone and intra-layer small-world connectivity. Layer $n=1,\ldots,N$ contains
\begin{equation}
    Q_n=m^{n-1}
\end{equation}
vertices, so that
\begin{equation}
    S_m^N=\sum_{n=1}^{N}Q_n=\frac{m^N-1}{m-1}.
\end{equation}
Each vertex in layer $n$ is connected to $m$ vertices in layer $n+1$, defining the inter-layer edge set $E_{\rm tree}$. Within each layer, the vertices are first connected to their $K$ nearest neighbours under periodic boundary conditions and are then rewired with probability $\phi$ according to the Watts--Strogatz prescription, excluding self-loops and repeated edges. The full edge set is
\begin{equation}
    E=E_{\rm tree}\cup\left(\bigcup_{n=1}^N E_n^{\rm SW}\right).
\end{equation}
When $K>Q_n-1$, the corresponding layer is complete and no additional intra-layer edges are added.

For each graph realization, the adjacency matrix is $A$, the degree matrix is $D$, and the graph Laplacian is $L=D-A$. The continuous-time quantum-walk Hamiltonian is
\begin{equation}
    H=-\gamma L=-\sum_i \omega_i\ketbra{i}{i}+\sum_{i,j}t_{ij}\ketbra{i}{j},
\end{equation}
with $\omega_i=\gamma d_i$ and $t_{ij}=\gamma A_{ij}$. We set $\gamma=1$ throughout. Observables are averaged over $R$ graph/percolation realizations according to
\begin{equation}
    \overline{O}=\frac{1}{R}\sum_{r=1}^R O_r.
\end{equation}

\subsection{Bond percolation and geometrical threshold estimate}

Structural disorder is introduced by independent bond percolation. Each edge is retained with probability $p$ and removed with probability $1-p$, so that the hopping amplitudes in each percolated realization are randomized as $t_{ij}=1$ with probability $p$ and $t_{ij}=0$ with probability $1-p$. The dilution modifies both the off-diagonal hopping structure and the diagonal degree-dependent terms.

For a pure $m$-branching tree, the classical bond-percolation threshold is
\begin{equation}
    p_c^{\rm tree}=\frac{1}{m}.
\end{equation}
Intra-layer links lower the effective geometrical scale by creating alternative routes across a layer before downward propagation. In the regular intra-layer limit $\phi=0$, let $\mathcal P_n$ be the probability of reaching layer $n$. Retaining neighbour shells up to range $R$, one obtains the recurrence
\begin{equation}
\begin{split}
    \mathcal{P}_n
    =1
    &-
    \left(1-p\mathcal{P}_{n-1}\right)^m \\
    &\times
    \prod_{r=1}^{R}
    \left\{
    1-p^r
    \left[
    1-\left(1-p\mathcal{P}_{n-1}\right)^m
    \right]
    \right\}^{K^r}.
\end{split}
\end{equation}
Writing the right-hand side as $f(\mathcal P_{n-1},p)$, the onset of a nonzero reaching probability is estimated from $\partial_x f(x,p)|_{x=0}>1$, giving
\begin{equation}
    p_c\simeq\frac{1}{m+K}.
\end{equation}
This estimate is used only as a geometrical scale, not as an exact critical threshold for the full rewired architecture.

\subsection{Spectral structure of percolated realizations}

\begin{figure*}[t]
    \centering
    \includegraphics[width=\linewidth]{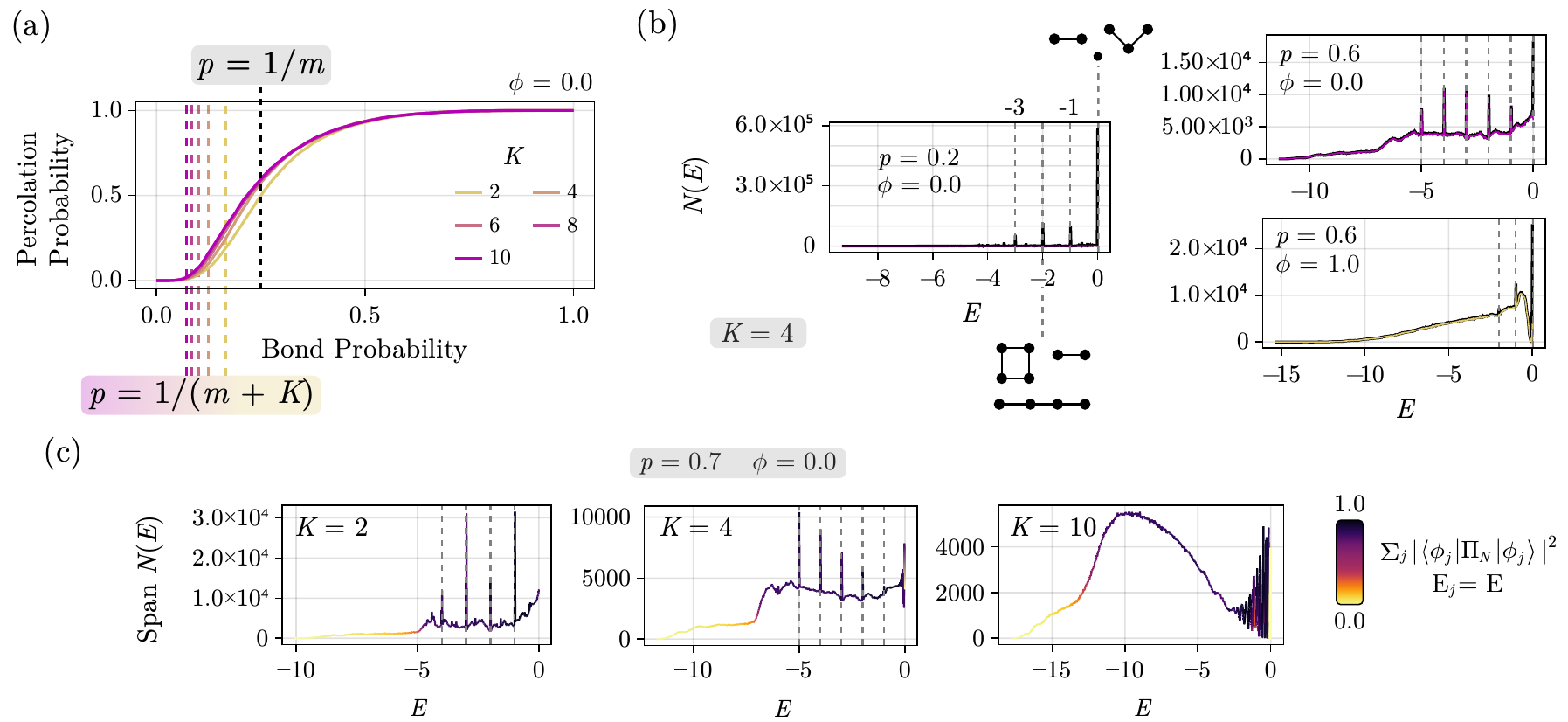}
    \caption{Percolation and spectral structure. (a) Percolation probability as a function of bond probability $p$ for $K=2,4,6,8,10$ and $\phi=0$. Dashed lines mark $p=1/m$ and $p=1/(m+K)$. (b) Number of states $N(E)$ for representative bond probabilities and rewiring regimes. Repeated local motifs produce sharp spectral peaks. (c) Spectral weight associated with the spanning cluster for different $K$, with the colour scale indicating eigenstate overlap with the final layer.}
    \label{fig:spectral}
\end{figure*}

Figure~\ref{fig:spectral} shows the percolation probability and spectral properties used to interpret the coherent transport optimum. At small $p$, the graph decomposes into repeated finite motifs, producing sharp peaks in the density of states. Increasing $p$ creates larger connected components and increases the spectral weight associated with the spanning cluster. Rewiring smooths part of the motif-induced peak structure by reducing repeated local patterns.

The final-layer projector is
\begin{equation}
    \Pi_N=\sum_{i\in N}\ketbra{i}{i}.
\end{equation}
Eigenstates with simultaneous overlap with the root and with the final-layer subspace are the states that contribute most directly to root-to-boundary transport. The long-time coherent efficiency is therefore sensitive to the spectral organization of the percolated Hamiltonian, not only to geometrical connectivity.

\subsection{Layer-resolved coherent dynamics}

\begin{figure}[t]
    \centering
    \includegraphics[width=\linewidth]{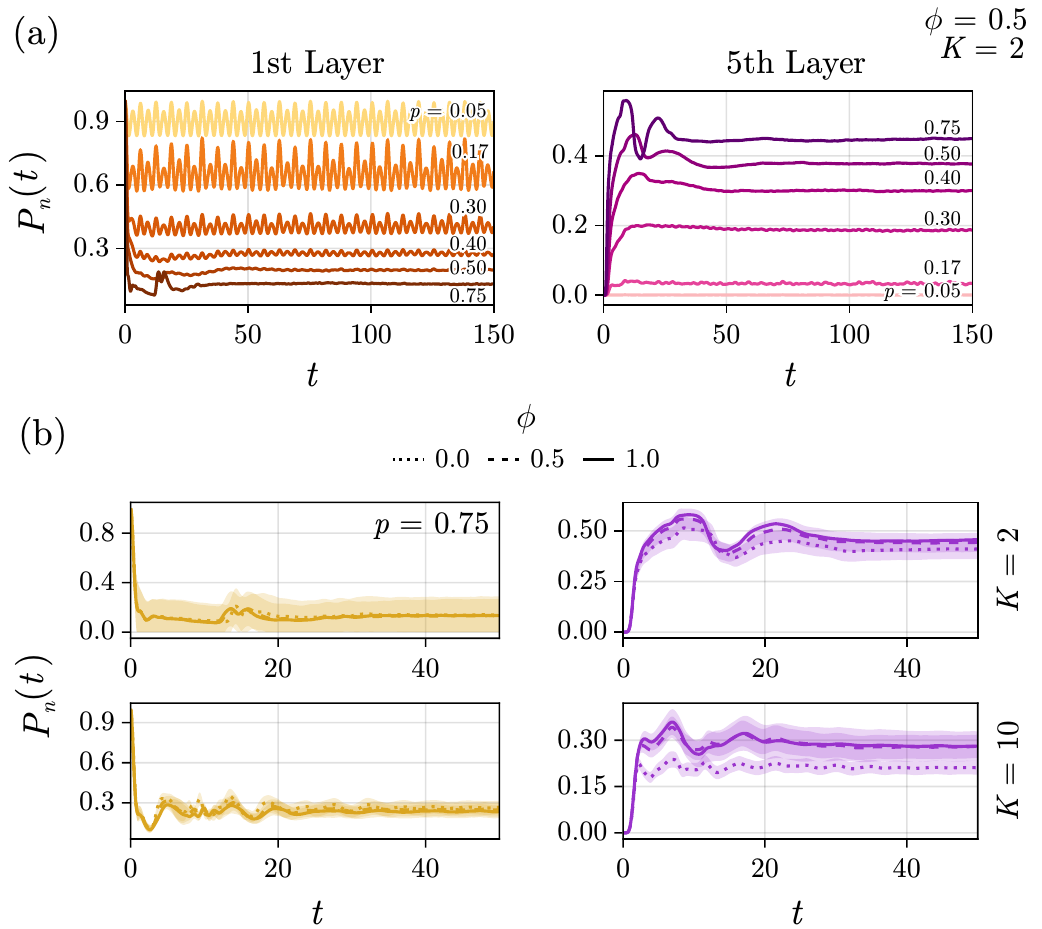}
    \caption{Layer-resolved coherent dynamics. (a) First-layer and final-layer probabilities for different bond probabilities at fixed $(\phi,K)=(0.5,2)$. (b) Effect of rewiring probability for $K=2$ and $K=10$ at fixed $p=0.75$. Shaded regions indicate ensemble fluctuations over $R=10000$ realizations.}
    \label{fig:dynamics}
\end{figure}

The state evolves as
\begin{equation}
    \ket{\psi_r(t)}=e^{-iH_rt}\ket{1},
\end{equation}
with $H_r$ the Hamiltonian of the $r$th percolated realization. The layer-resolved probability is
\begin{equation}
    P_n(t)=\overline{\bra{\psi_r(t)}\Pi_n\ket{\psi_r(t)}}.
\end{equation}
Figure~\ref{fig:dynamics} shows representative time traces. At low $p$, the excitation remains near the root and $P_N(t)$ is small. Increasing $p$ opens inter-layer and intra-layer paths and increases final-layer population. At larger connectivity, ensemble-averaged oscillations are damped by averaging over distinct coherent spectra.

\subsection{Dephased and classical benchmark}

To compare coherent transport with an incoherent benchmark, we use site-basis dephasing,
\begin{equation}
    \dot\rho_r(t)=-i[H_r,\rho_r(t)]-q\left[\rho_r(t)-\sum_jP_j\rho_r(t)P_j\right],
\end{equation}
where $P_j=\ketbra{j}{j}$. The layer probability is
\begin{equation}
    P_n^{\rm deph}(t)=\overline{\mathrm{Tr}[\Pi_n\rho_r(t)]}.
\end{equation}
The dephased dynamics suppresses the phase-sensitive interference responsible for the coherent optimum and provides an operational benchmark for distinguishing coherent transport effects from geometrical percolation alone. In the main text we compare the coherent final-layer efficiency with the corresponding dephased/classical layer probabilities.

Depending on the chosen classical generator, one may use either the combinatorial-Laplacian reference or a normalized random-walk reference. For a random-walk generator on each connected component $\mathcal C_r$, the stationary distribution is degree-weighted,
\begin{equation}
    \pi_j^{(r)}=\frac{d_j^{(r)}}{2|E[\mathcal C_r]|},\qquad j\in\mathcal C_r,
\end{equation}
where $|E[\mathcal C_r]|$ is the number of edges in the connected component. The corresponding layer probability is
\begin{equation}
    \chi_n^{\rm cl}=\overline{\sum_{j\in n\cap\mathcal C_r}\pi_j^{(r)}}.
\end{equation}
This is the benchmark used for the classical contrast plotted in the main text. If instead the combinatorial Laplacian $e^{-Lt}$ is used directly as a classical generator, the stationary distribution is uniform on each connected component. The choice of benchmark should therefore be stated together with the generator.

\subsection{Additional remarks on applications}

The main text focuses on the transport principle. Possible applications include coherent photonic networks, where redundant links can improve robustness but also introduce interference pathways, and biologically inspired hierarchical transport models, where lateral connectivity and structural disorder coexist. Direct biological applications require mechanisms beyond the present tight-binding model, such as non-Hermitian losses, non-Markovian memory, or environment-assisted transport.

\end{document}